\documentclass[10pt,a4paper,twocolumn,english,aps,manuscript,superscriptaddress, aps,manuscript,preprint,showpacs]{revtex4}

\usepackage[T1]{fontenc}
\usepackage[latin9]{inputenc}
\usepackage{color}
\usepackage{babel}
\usepackage{float}
\usepackage{amstext}
\usepackage{amssymb}
\usepackage{graphicx}
\usepackage{esint}
\usepackage[unicode=true,pdfusetitle,
 bookmarks=true,bookmarksnumbered=true,bookmarksopen=false,
 breaklinks=true,pdfborder={0 0 0},pdfborderstyle={},backref=false,colorlinks=true]
 {hyperref}
\hypersetup{
 linkcolor=red, citecolor=blue,  urlcolor=blue}

\makeatletter

\newcommand{\lyxmathsym}[1]{\ifmmode\begingroup\def\b@ld{bold}
  \text{\ifx\math@version\b@ld\bfseries\fi#1}\endgroup\else#1\fi}

\providecommand{\tabularnewline}{\\}

\@ifundefined{textcolor}{}
{%
 \definecolor{BLACK}{gray}{0}
 \definecolor{WHITE}{gray}{1}
 \definecolor{RED}{rgb}{1,0,0}
 \definecolor{GREEN}{rgb}{0,1,0}
 \definecolor{BLUE}{rgb}{0,0,1}
 \definecolor{CYAN}{cmyk}{1,0,0,0}
 \definecolor{MAGENTA}{cmyk}{0,1,0,0}
 \definecolor{YELLOW}{cmyk}{0,0,1,0}
}

\makeatother

\makeatother

\begin{document}
\title{Structural and magnetic properties of a new cubic spinel LiRhMn$\mathbf{O}_{4}$}
\author{S. Kundu}
\email{skundu37@gmail.com}

\affiliation{Department of Physics, Indian Institute of Technology Bombay, Powai,
Mumbai 400076, India}
\author{T. Dey}
\affiliation{Experimental Physics VI, Center for Electronic Correlations and Magnetism,
University of Augsburg, 86159 Augsburg, Germany}
\author{M. Prinz-Zwick}
\affiliation{Experimental Physics V, Center for Electronic Correlations and Magnetism,
University of Augsburg, 86159 Augsburg, Germany}
\author{N. $\mathrm{B\mathrm{\ddot{u}}ttgen}$}
\affiliation{Experimental Physics V, Center for Electronic Correlations and Magnetism,
University of Augsburg, 86159 Augsburg, Germany}
\author{A. V. Mahajan}
\email{mahajan@phy.iitb.ac.in}

\affiliation{Department of Physics, Indian Institute of Technology Bombay, Powai,
Mumbai 400076, India}
\date{\today}
\begin{abstract}
We report the structural and magnetic properties of a new system LiRhMnO$_{4}$ (LRMO) through x-ray diffraction, bulk magnetization, heat capacity and $^{7}$Li nuclear magnetic resonance (NMR) measurements. LRMO
crystallizes in the cubic space group $\mathit{Fd}$$\bar{3}$$\mathit{m}$.
From the DC susceptibility data, we obtained the Curie-Weiss temperature
$\mathrm{\theta}_{\mathrm{CW}}$ = -26\,K and Curie constant $\mathit{C}$
= 1.79\,$\mathrm{{\normalcolor Kcm^{3}/mol}}$ suggesting antiferromagnetic
correlations among the magnetic Mn$^{4+}$ ions with an effective
spin $\mathit{S}$ = $\frac{3}{2}$. At $\mathit{H}$ = 50 Oe, the
field cooled and zero-field cooled magnetizations bifurcate at a freezing
temperature, $T_{f}$ = 4.45\,K, which yields the frustration parameter
$\mathit{f=\frac{\mid\theta_{CW}\mid}{T_{f}}}>5$. AC susceptibility,
shows a cusp-like peak at around $T_{f}$, with the peak position
shifting as a function of the driving frequency, confirming a spin-glass-like
transition in LRMO. LRMO also shows typical spin-glass characteristics
such as memory effect, aging effect and relaxation. In the heat capacity,
there is no sharp anomaly down to 2\,K indicative of long-range ordering.
The field sweep $^{7}$Li NMR spectra show broadening with decreasing
temperature without any spectral line shift. The $^{7}$Li NMR spin-lattice
and spin-spin relaxation rates also show anomalies due to spin freezing
near $T_{f}$.
\end{abstract}
\keywords{Geometric frustration, Spinel, Spin-glass, Memory effect, Aging effect,
NMR.}
\pacs{75.50.Lk, 75.40.Cx, 76.60.-k}
\maketitle

\section{introduction}

In the last few decades, most of the scientific work in condensed
matter physics has chiefly been devoted to study the strongly correlated
electron systems (SCES) \cite{Dagotto2005,Dagotto2008}. Materials
with strong electronic correlations are the materials, in which the
movement of one electron depends on the positions and movements of
all other electrons due to the long-range Coulomb interaction (U).
In this regard, the transition metal oxide (TMO) compounds \cite{Tokura2000}
have become the centre stage of attraction to physicists, since the
TMO have outermost electrons in d-orbitals which are strongly localized.
Hence, the electron density is no longer homogeneous and the striking
properties of the system are in fact dependent on the presence of
strong electron-electron interactions. Also, the frustration in TMO,
either imposed by the geometry of the spin system or by the competing
interactions, leads to exotic behavior \cite{Ramirez1994,Greedan2001,Moessner2006,Balents2010}.
The rich physics of magnetically frustrated systems, continues to
attract interest in the condensed matter research community. The resurgence
of interest began with the discovery of high-$\mathit{T_{\mathrm{c}}}$
superconductivity (observed in layered cuprates \cite{Bednorz1986}
in the late 80's and pnictides \cite{Kamihara2008} more recently),
and novel phenomenon such as metal-insulator transition (MIT) \cite{Mott1968},
colossal magneto-resistance (CMR) \cite{Tokura2000a}, charge ordering
and quantum magnetism. It was soon realized that the strong interplay
of spin, charge, lattice and orbital degrees of freedom in these correlated
systems resulted in such diverse properties. In recent years, the
interest in solids containing lithium ions has increased profoundly
due to the potential applications in long-lasting rechargeable batteries.
In this regard, the spinel oxide LiMn$_{2}$O$_{4}$ has attracted
wide attention as a cathode material of batteries due to its low cost
and non-toxicity \cite{Eftekhari2003,Arillo2005}. At room temperature
LiMn$_{2}$O$_{4}$ \cite{Thackeray1983} crystallizes in the cubic
space group $\mathit{Fd}$$\bar{3}$$\mathit{m}$, with the following
cation distribution (Li$^{+}$)$_{\mathrm{A}}${[}Mn$^{3+}$Mn$^{4+}${]}$_{\mathrm{B}}$O$_{4}$;
here the subscripts A and B denote the tetrahedral and octahedral
sites, respectively. Likewise, Takagi \textit{et al.} found the metal-insulator
transition (MIT) property in $\mathrm{LiRh_{2}O_{4}}$\cite{Okamoto2008,Arita2008,Knox2013},
which behaves like a paramagnetic metal at high temperature; whereas
below about 170\,K it becomes a valence bond insulator and the ground
state of mixed-valent $\mathrm{Li[Rh^{3+}(\mathit{S}=0)Rh^{4+}(\mathit{S}=\frac{1}{2})]O_{4}}$
is charge frustrated. How will the ground state of this system vary
if one replaces the $\mathit{S}$ = $\frac{1}{2}$ ion with higher
spin ions, say $\mathit{S}$ = $\frac{3}{2}$ ? With this motivation,
we decided to explore $\mathrm{LiRhMnO_{4}}$ (LRMO) which is structurally
identical to $\mathrm{LiMn_{2}O_{4}}$ but magnetically different.
Magnetic properties of LRMO have so far not been reported. Only the
structure of LRMO was first reported long back in 1963 by G. Blasse
\cite{Blasse1963}. It is a mixed metal oxide with spinel structure
\cite{Onoda1997} where 50\% of the B-sites are occupied by non-magnetic
$\mathrm{Rh^{3+}}$($\mathit{S}$ = 0) and the other 50\% by magnetic
$\mathrm{Mn^{4+}}$($\mathit{S}$ = $\frac{3}{2}$) ions. Usually,
the B-site spinel has a corner-shared tetrahedral network like the
pyrochlore lattice which is geometrically frustrated. But due to the
B-site disorder, the frustration may be relieved and result in a spin-glass
state \cite{Mydosh1993} at a low $\mathit{T}$.

We have synthesized polycrystalline LRMO and studied its bulk and
local magnetic properties through various characterization techniques
such as x-ray diffraction, DC and AC magnetization, heat capacity
and field sweep $^{7}$Li nuclear magnetic resonance (NMR). We found
that LRMO has antiferromagnetic (AFM) correlations among Mn$^{4+}$
ions and conventional spin-glass ground state with the spin-freezing
temperature $\mathit{T_{f}}$ = 4.45\,K.

\section{experimental details}

Polycrystalline ${\normalcolor \mathrm{LRMO}}$ was prepared by solid-state
reaction. Pre-heated starting materials (Li$_{2}$CO$_{3}$, Rh metal
powder and MnO$_{2})$ were mixed in stoichiometry and ground thoroughly
for hours. Finally a hard pellet was made and calcined at ${\normalcolor 500^{\mathrm{o}}\mathrm{C}}$,
${\normalcolor \mathrm{700^{o}C}}$, ${\normalcolor \mathrm{900^{o}C}}$
and ${\normalcolor \mathrm{1050^{o}C}}$ for 24 hours each time. As
there is a chance of lithium evaporation above 900$^{\mathrm{o}}$C,
15\% excess Li$_{2}$CO$_{3}$ was mixed to get the pure LRMO. The
processes of grinding and firing were done until we obtained the single
phase sample. The single phase of LRMO is confirmed from the powder
x-ray diffraction (XRD) measurements at room temperature with Cu $K_{\alpha}$
radiation ($\lambda=1.54182\,\lyxmathsym{\AA}$) on a PANalytical
X\textquoteright Pert PRO diffractometer. DC and AC magnetization
data were measured as a function of temperature $\mathit{T}$ $(2\thinspace\mathrm{K}-400$~\,K)
with the applied field $\mathit{H}$ $(0-70$\,kOe) and the frequency
$\nu$ (1$\thinspace\mathrm{Hz}-$1000\,Hz) using a commercial superconducting
quantum interference device (SQUID) magnetometer. Low-field magnetization
measurements were performed utilizing the reset magnet option
of the SQUID. Heat capacity measurements were performed in the temperature
$\mathit{T}$ $(2\thinspace\mathrm{K}-215$\,K) and in the field
$\mathit{H}$ $(0-90$\,kOe) using a Quantum Design PPMS. As the $^{7}$Li NMR spectra are very broad especially
at low-$\mathit{T}$ and it is difficult to obtain the full line-shape
only by the Fourier transform of the time echo signal in our fixed
field NMR setup, we have performed field sweep $^{7}$Li NMR measurements
at 60\,MHz and 95\,MHz. The spin-lattice relaxation rate ($\frac{1}{T_{1}}$)
is measured by the saturation recovery method and the spin-spin relaxation
rate ($\frac{1}{T_{2}}$) is obtained by measuring the decay of the
echo integral with variable delay time.

\begin{table}[H]
\caption{\label{tab:Atomic-Co-ordinates-of LRMO}{\small{}Atomic coordinates
of $\mathrm{LRMO}$}}
\vspace{0.5cm}

\centering{}%
\begin{tabular}{cccccc}
\hline 
Atom & Wyckoff position & x & y & z & Occupancy\tabularnewline
\hline 
\hline 
Li & 8a & 0.125 & 0.125 & 0.125 & 1.00\tabularnewline
Rh & 16d & 0.500 & 0.500 & 0.500 & 0.50\tabularnewline
Mn & 16d & 0.500 & 0.500 & 0.500 & 0.50\tabularnewline
O & 32e & 0.747 & 0.747 & 0.747 & 1.00\tabularnewline
\hline 
\end{tabular}
\end{table}

\section{Results and Discussion}
\begin{center}
\textbf{\large{}A. Crystal structure}{\large\par}
\par\end{center}

\begin{figure}[H]
\centering{}\includegraphics[scale=0.33]{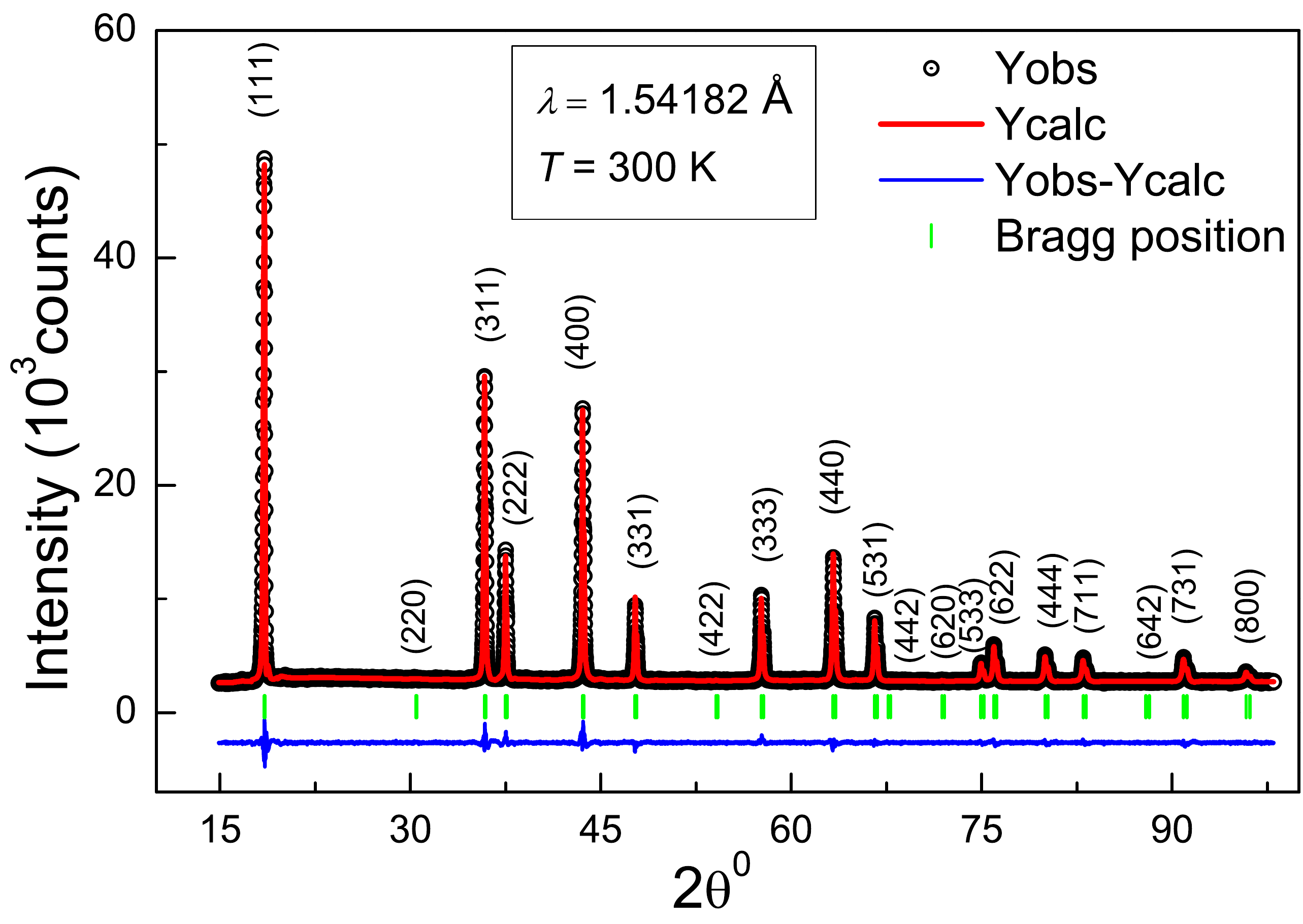}\caption{\label{fig:Refinment-of LRMO}{\small{}The Rietveld refinement of
room temperature powder XRD pattern of $\mathrm{LRMO}$ is shown along
with its Bragg peak positions (green vertical marks) and the corresponding
Miller indices (hkl).}}
\end{figure}

The powder XRD data has been recorded with Cu-$K_{\alpha}$ radiation
over the angular range $\mathrm{15^{0}\leqslant2\theta\leqslant98^{0}}$
in $\mathrm{0.004}{}^{0}$ step size and treated by profile analysis
using the Rietveld refinement \cite{Rietveld1969} by Fullprof suite
\cite{Rodriguez-Carvajal1993} program. From the XRD pattern analysis,
we found that the prepared \textcolor{black}{$\mathrm{LRMO}$ is crystallized
in single phase }and there is no sign of any unreacted ingredients
or impurity phases. The Rietveld refinement of XRD pattern is shown
in Fig. \ref{fig:Refinment-of LRMO}. From refinement, we obtained
the cell parameters of $\mathrm{LRMO}$, $\mathit{a}$ = $\mathit{b}$
= $\mathit{c}$ = 8.319\,Å (which is close to the earlier reported
value 8.30\,Å \cite{Blasse1963}), $\alpha$=$\beta$= $\gamma=$$\mathrm{90^{o}}$
and the atomic coordinates of LRMO is given in Table \ref{tab:Atomic-Co-ordinates-of LRMO}.
The reliability of the x-ray refinement of LRMO is given by the following
parameters $\mathrm{\chi}^{2}$: 4.63; $\mathrm{\mathit{R}}_{\mathrm{p}}$:
2.98\%; $\mathrm{\mathit{R}}_{\mathrm{wp}}$: 5.68\%; $\mathrm{\mathit{R}}_{\mathrm{exp}}$:
2.63\%. 

\begin{figure}[h]
\centering{}\includegraphics[scale=0.38]{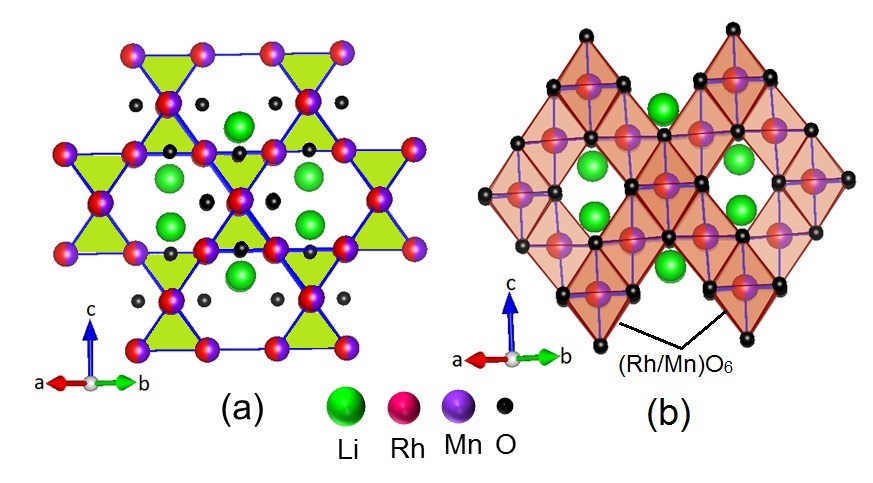}\caption{\label{fig:Structure-of-LRMO}{\small{}Structure of LRMO (a) The corner
shared tetrahedral network of (Rh/Mn) atoms in 3-dimension. (b) one
unit cell is shown with the edge-shared $\mathrm{(Rh/Mn)O_{6}}$ octahedra.}}
\end{figure}
The structure of LRMO has been drawn and analyzed by using Vesta software
\cite{Momma2011}. We have obtained the atomic coordinates from Rietveld
refinement done on XRD pattern of LRMO which crystallizes in the non-centrosymmetric
cubic spinel structure $\mathit{Fd}$$\bar{3}$$\mathit{m}$ (space
group 227). The Rh or Mn atoms are connected to each other via a tetrahedral
network as shown in Fig. \ref{fig:Structure-of-LRMO}(a). These tetrahedral
are corner-shared and form a geometrically frustrated magnetic system.
In the structure, $\mathrm{(Rh/Mn)O_{6}}$ form perfect octahedra
with (Rh/Mn)-O bond distance 2.055\,Å (shown in Fig. \ref{fig:Structure-of-LRMO}(b)).
The presence of non-magnetic $\mathrm{Rh^{3+}}$($\mathit{S}$ = 0)
at the B-site of the spinel, $\mathit{i.e.}$ in a tetrahedral unit,
distorts the corner-shared arrangement of $\mathrm{Mn^{4+}}$($\mathit{S}$
= $\frac{3}{2}$) ions. This makes the B-sites diluted.
\begin{center}
\textbf{\large{}B. Bulk} \textbf{\large{}magnetization}{\large\par}
\par\end{center}

\begin{center}
\textbf{1. DC susceptibility}
\par\end{center}

\begin{figure}[h]
\centering{}\includegraphics[scale=0.35]{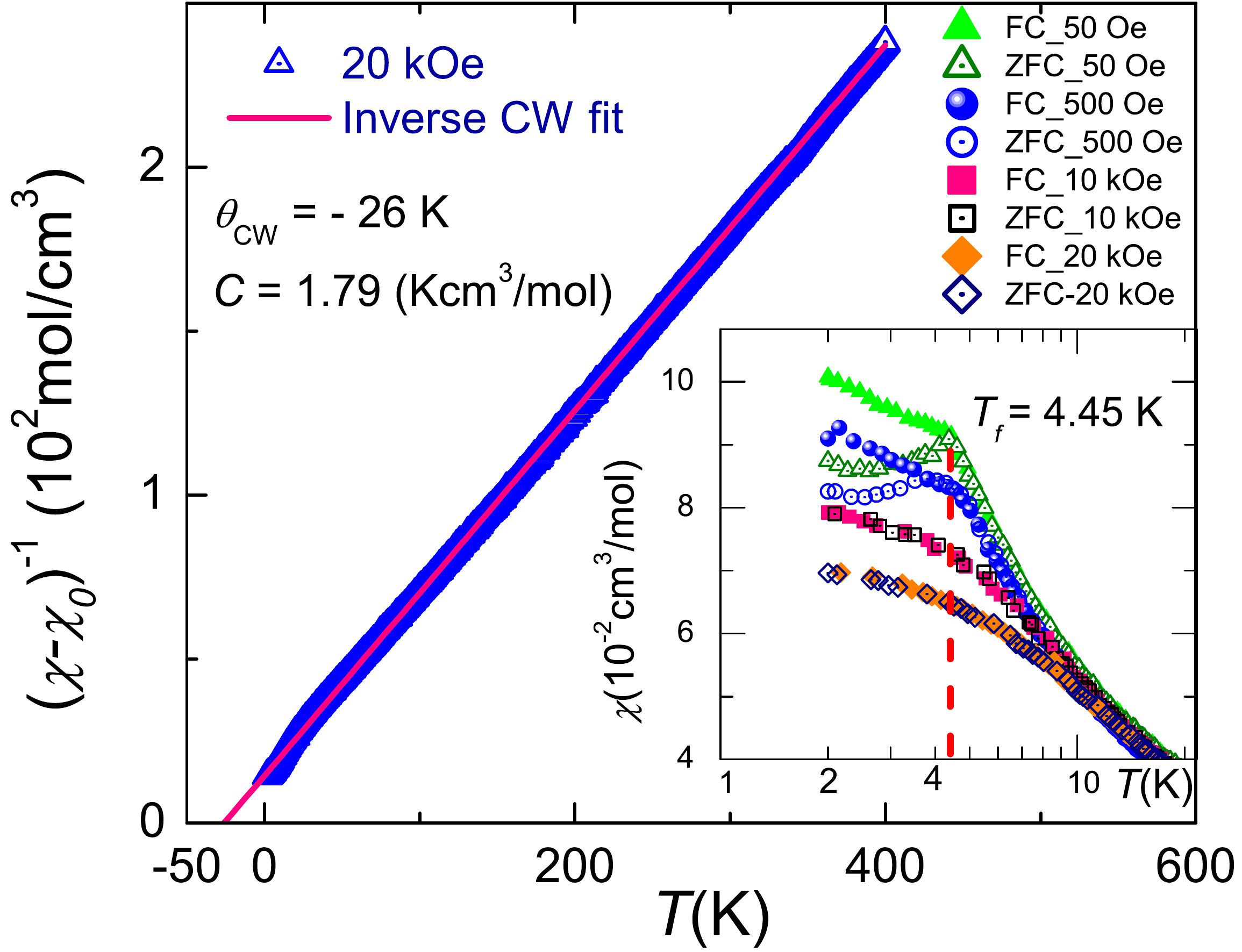}\caption{\label{fig:Temperature-dependence-of LRMO}{\small{}The temperature
dependence of $\chi(T)$ of LRMO at $\mathit{H}$ = 20\,kOe. The
inset, (semi-log scale) shows the ZFC-FC bifurcation below 4.45\,K
at different applied fields.}}
\end{figure}

The temperature dependence of the bulk dc magnetic susceptibility
$\chi(T)=\frac{M(T)}{H}$ is measured on LRMO under different applied
magnetic fields in the temperature range of (2-400)\,K. The main
features of our observations from the dc susceptibility measurement
are discussed here. With increasing fields, the $\chi(T)$ reduces
in the low temperature region (see inset of Fig. \ref{fig:Temperature-dependence-of LRMO}).
Below 5\,K, there is splitting between the zero-field cooled (ZFC)
and field cooled (FC) data at $\mathit{H}$ = 50\,Oe and 500\,Oe
as shown in the inset of Fig. \ref{fig:Temperature-dependence-of LRMO}.
Also, the $\chi(T)$ below 500\,Oe shows some anomaly around 5\,K.
This may be due to regular antiferromagnetic (AFM) ordering which
is very sensitive to the applied field as splitting between ZFC-FC
is suppressed with fields higher than 5\,kOe. The existence of ZFC-FC
splitting below 500\,Oe suggests the presence of a glassy state below
5\,K. This is a signature of conventional spin-glass \cite{Binder1986}.
Fig. \ref{fig:Temperature-dependence-of LRMO} shows the paramagnetic
behavior of the dc susceptibility at 20\,kOe. The Curie-Weiss fitting
in the high temperature region (200-400\,K) gives a Curie constant
$\mathit{C}$ = 1.79\,$\mathrm{{\normalcolor Kcm^{3}/mol}}$ and
a Curie-Weiss temperature $\theta_{\mathrm{CW}}=-26\mathrm{\thinspace K}$.
The negative value of the Curie-Weiss temperature suggests AFM interaction
among the magnetic $\mathrm{Mn^{4+}}$ ions. The effective moment
of $\mathrm{\mathrm{M}n^{4+}}$ ions {[}using $\mathit{C}$ = 1.79\,$\mathrm{{\normalcolor Kcm^{3}/mol}}${]}
is $\mu_{\mathrm{eff}}=3.78{\normalcolor \,\mu_{\mathrm{B}}}$ which
is close to the expected value 3.87\,$\mu_{\mathrm{B}}$ for the
$\mathrm{Mn^{4+}}$ ion.

\begin{figure}[h]
\centering{}\includegraphics[scale=0.35]{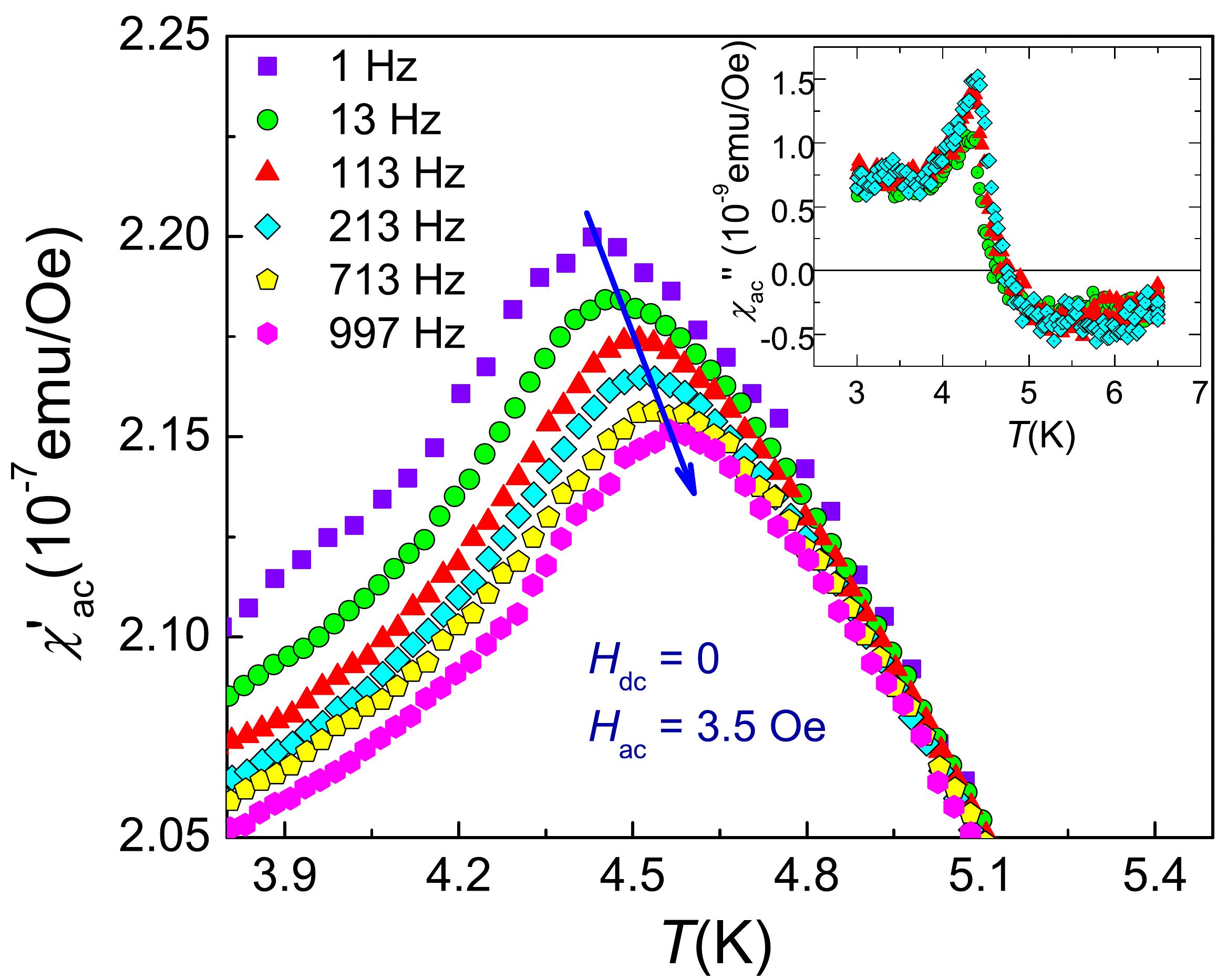}\caption{\label{fig:AC-susceptibility_LRMO} {\small{}The in-phase component
of ac susceptibility $\chi_{\mathrm{ac}}^{'}(T)$ of LRMO as a function
of temperature at different frequencies are shown in the main figure.
In the inset, $\chi_{\mathrm{ac}}^{''}(T)$ is shown.}}
\end{figure}

\begin{center}
\textbf{2. AC susceptibility}
\par\end{center}

The ac susceptibility is measured by keeping the dc applied field
to be zero and with an ac field $H_{\mathrm{ac}}$ of 3.5\,Oe amplitude.
The frequency dependence of the in-phase component $\chi_{\mathrm{ac}}^{'}(T)$
is shown in Fig. \ref{fig:AC-susceptibility_LRMO}. The freezing temperature
($\mathrm{\mathit{T}}_{f}$) shifts towards higher temperatures as
the frequency increases which are typical features in glassy systems
\cite{Mulder1981}. Also, the out of phase component of the ac susceptibility
$\chi_{\mathrm{ac}}^{''}(T)$, shown in the inset of Fig. \ref{fig:AC-susceptibility_LRMO}, has a frequency dependence with an anomaly
around $T_{f}$. Below and above $T_{f}$, the value of $\chi_{\mathrm{ac}}^{''}(T)$ is non-zero positive and negative respectively. These observations confirm a spin-glass ground state of LRMO. Usually, the frequency dependence of spin-freezing temperature ($T_{f}$) is estimated in terms of the relative shift ( $\delta T_{f}$) of the $T_{f}$, defined as $\delta T_{f}$ = $ \frac {\Delta T_{f}} {T_{f} \mathrm{\Delta log}_{10}(\nu)}$ \cite{Mahendiran2003}, which is found to be 0.022 for LRMO. It is interesting to note that the value of $\delta T_{f}$ which determines the sensitivity to the frequency, falls in between the value of conventional spin-glasses and superparamagnets. The present value of $\delta T_{f}$ is close to 0.037 which is observed for metallic glasses \cite{Luo2008}.

\begin{figure}[h]
\centering{}\includegraphics[scale=0.37]{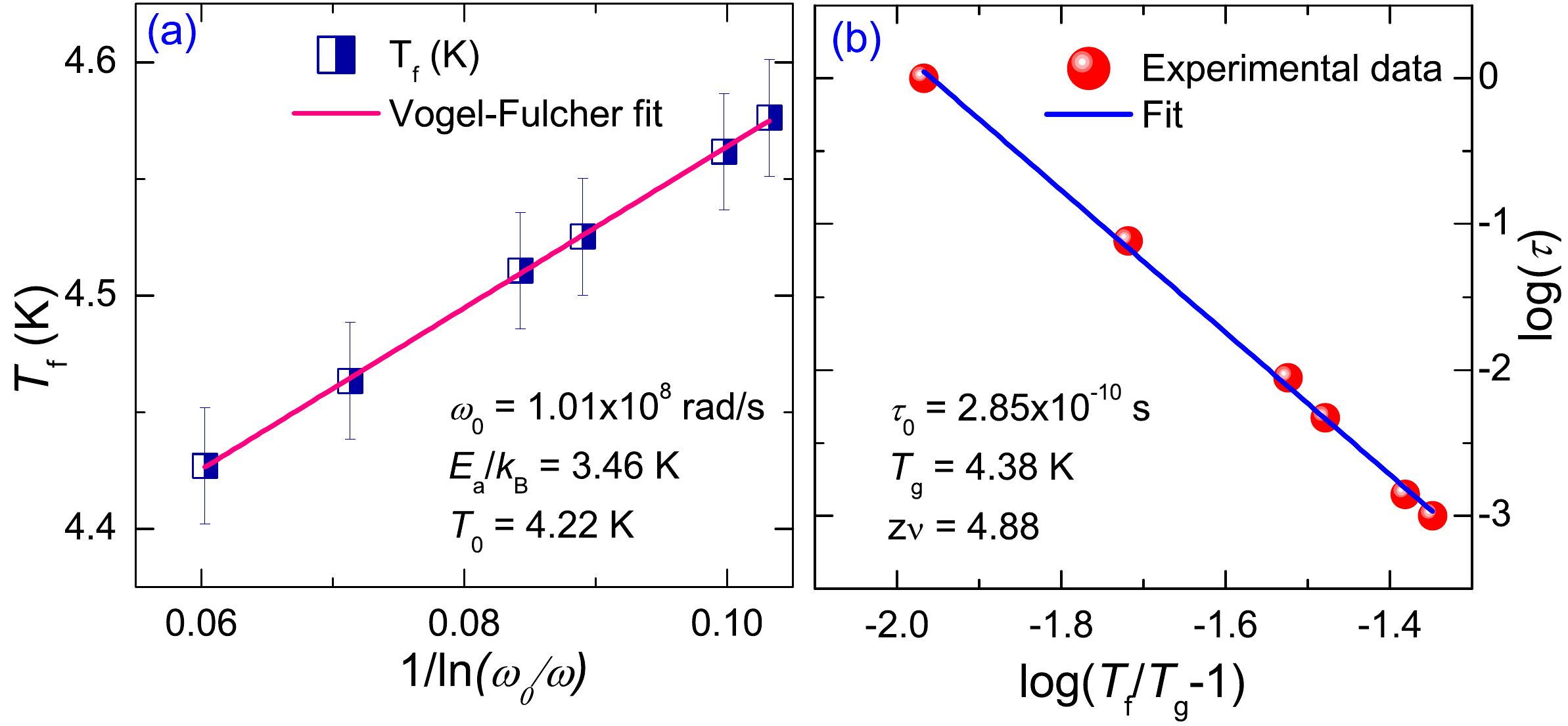}\caption{\label{fig:Vogel-Fulcher-fit-of LRMO}{\small{}(a) The Vogel-Fulcher
fit of the freezing temperatures $\mathit{T_{f}}$ at different frequencies
of the LRMO sample. \label{fig:critical exonent fit LRMO} (b) A fit
of the $\mathit{T_{f}}$ by the critical slowing down equation.}}
\end{figure}

The Vogel-Fulcher fit by equation $\left(\omega=\omega_{0}\thinspace exp[-\mathit{E}_{a}/k_{B}(T_{f}-T_{0})]\right)$
of the variations of the freezing temperature ($\mathrm{\mathit{T}}_{f}$)
with frequency suggests short-range Ising spin-glass behavior \cite{Fisher1986}
(shown in Fig. \ref{fig:Vogel-Fulcher-fit-of LRMO}). From the fit,
we obtained the activation energy $\mathrm{\mathit{E}}_{a}$/ $k_{\mathrm{B}}$
$\approx$ 3.46\,K, the characteristic angular frequency $\omega_{0}\approx$
$1.01\times10^{8}$\,rad/s and the Vogel\textendash Fulcher temperature
$\mathit{T_{\mathrm{0}}\approx\mathrm{4.22}\mathrm{\thinspace K}}$.
For a conventional spin-glass system, $\mathit{\omega_{0}}$ is of
the order of $10^{13}$ \,rad/s. So the obtained $\mathit{\omega_{0}}$
value is small compared to that of a usual spin-glass system. This
large deviation may not be the true scenario as the error involved
in determining the freezing temperature is large and the measured
frequency range is limited to only two decades. 

The variation of the freezing temperature $T_{f}$ with frequency
obey with the critical slowing down dynamics (see Fig. \ref{fig:critical exonent fit LRMO})
which is expressed by the equation: $\tau$ = $\tau_{0}$( $T_{f}$
/ $T_{g}$ - 1)$^{-zv}$. Here $\tau_{0}$ is the relaxation time
and $\mathit{zv}$ is the dynamic exponent \cite{Dho2002}. We found
the best fit with $T_{g}$ = 4.38 K, $\tau_{0}$ $\approx$ 2.85 $\times$10$^{-10}$\,s
and $\mathit{zv}$ $\approx$ 4.88. The value of $\tau_{0}$ is 10$^{-10}$
to 10$^{-13}$\,s and $\mathit{zv}$ lies in between of 4 - 13 for
the conventional spin-glass systems \cite{Luo2008}. The present values
of $\tau_{0}$ and $\mathit{zv}$ imply that the ground state of LRMO
is a conventional spin-glass.

\begin{center}
\textbf{3. Memory effect}
\par\end{center}

\begin{figure}[h]
\centering{}\includegraphics[scale=0.35]{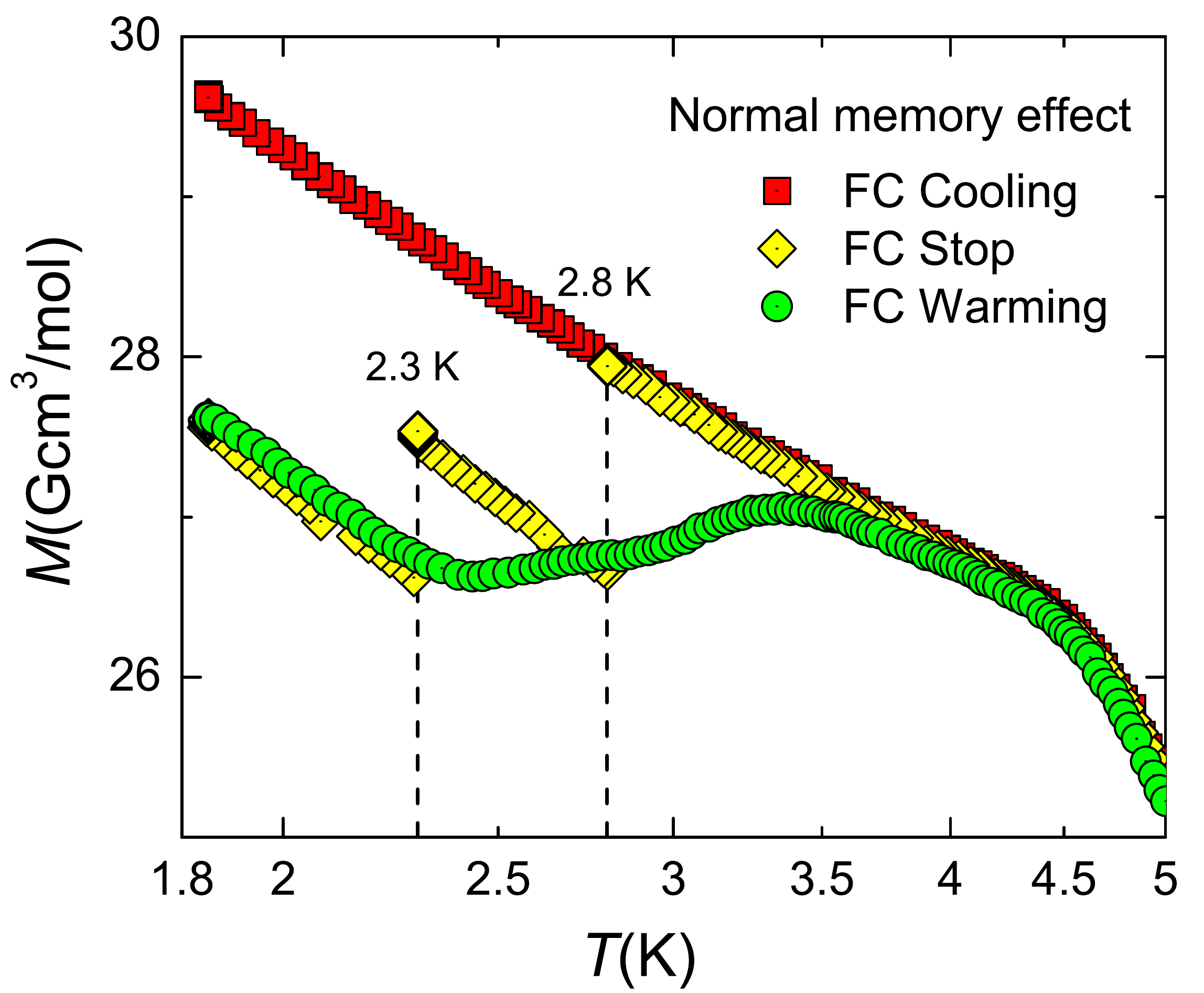}\caption{\label{fig:Memory-effect LRMO}{\small{}The memory effect in the field
cooled (FC) magnetization data of LRMO is plotted. The FC stop curve
is measured during the cooling of the LRMO sample in a field of 300\,Oe
with a relaxation at 2.8\,K and 2.3\,K for two hours, whereas the
FC warming curve is measured with continuous heating and the reference
curve, FC cooling, is measured by continuously cooling LRMO at $\mathit{H}$
= 300 Oe without any break.}}
\end{figure}

Fig. \ref{fig:Memory-effect LRMO} shows a memory effect in LRMO.
We have measured the field cooled (FC) magnetization using the protocol
described by Sun $\mathit{et\thinspace al.}$\cite{Sun2003,Chakrabarty2014}.
We have recorded the magnetization by cooling the LRMO sample down
to 1.85\,K with a cooling rate of 1\,K/min and an applied field
of 300\,Oe. We have interrupted the cooling process below $\mathit{T_{f}}$
i.e. at 2.8\,K and 2.3\,K for a waiting time $(t_{w})$ of 2\,hours.
We switched off the field during $\mathit{t_{w}}$, allowed the system
to relax and resumed the measurement after each stop and wait period.
Fig. \ref{fig:Memory-effect LRMO} shows step-like features which
are the evidence of stops at 2.8\,K and 2.3\,K in the FC stop curve.
Then we have recorded the magnetization of the sample while heating
continuously at the same field of 300 Oe. We also have measured a
reference curve named as ``FC cooling'' by simply cooling the sample
continuously at $\mathit{H}$ = 300\,Oe. We have noticed a change
of slope at 2.8\,K and a prominent minimum at 2.3\,K in the FC warming
curve. So it appears that the history of the sample magnetization is recorded as a memory in the sample. Such imprinting of memory has also been found in intermetallic systems GdCu \cite{Bhattacharyya2011},
Nd$_{5}$Ge$_{3}$\cite{Maji2011}) and in super spin-glass nanoparticle
system \cite{Sasaki2005,Sun2003}. This memory effect constitutes a standard observation in spin-glasses.

\begin{figure}[h]
\centering{}\includegraphics[scale=0.33]{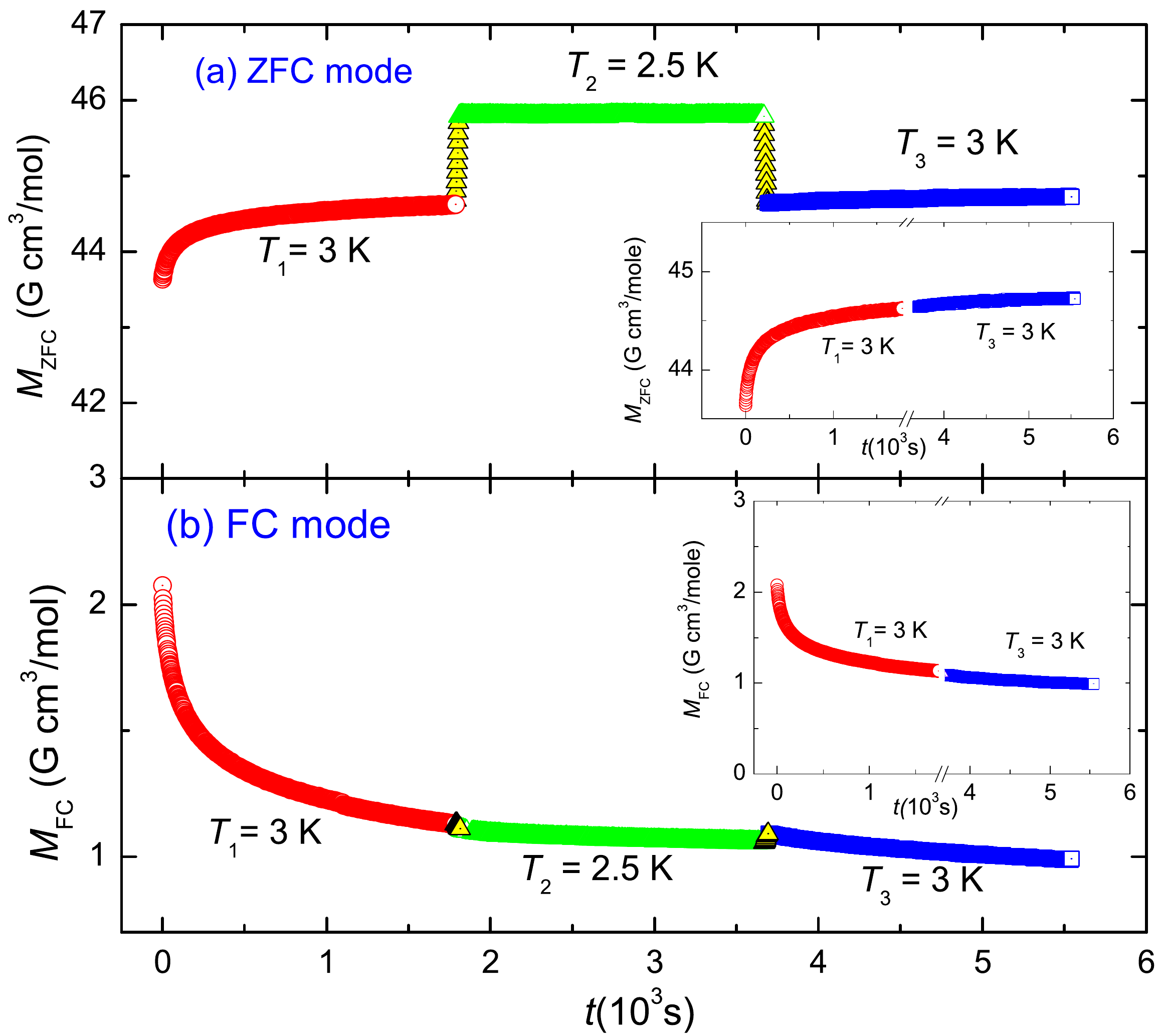}\caption{\label{fig:Negative cycle LRMO}{\small{}Magnetic relaxation in LRMO
at 3\,K with a temporary quench to 2.5\,K in (a) the ZFC mode and
(b) the FC mode. Insets show continuity in the relaxation data during
temperature $\mathrm{T}_{1}$ and $\mathrm{T}_{3}$ in both mode.}}
\end{figure}
\begin{figure}[h]
\centering{}\includegraphics[scale=0.33]{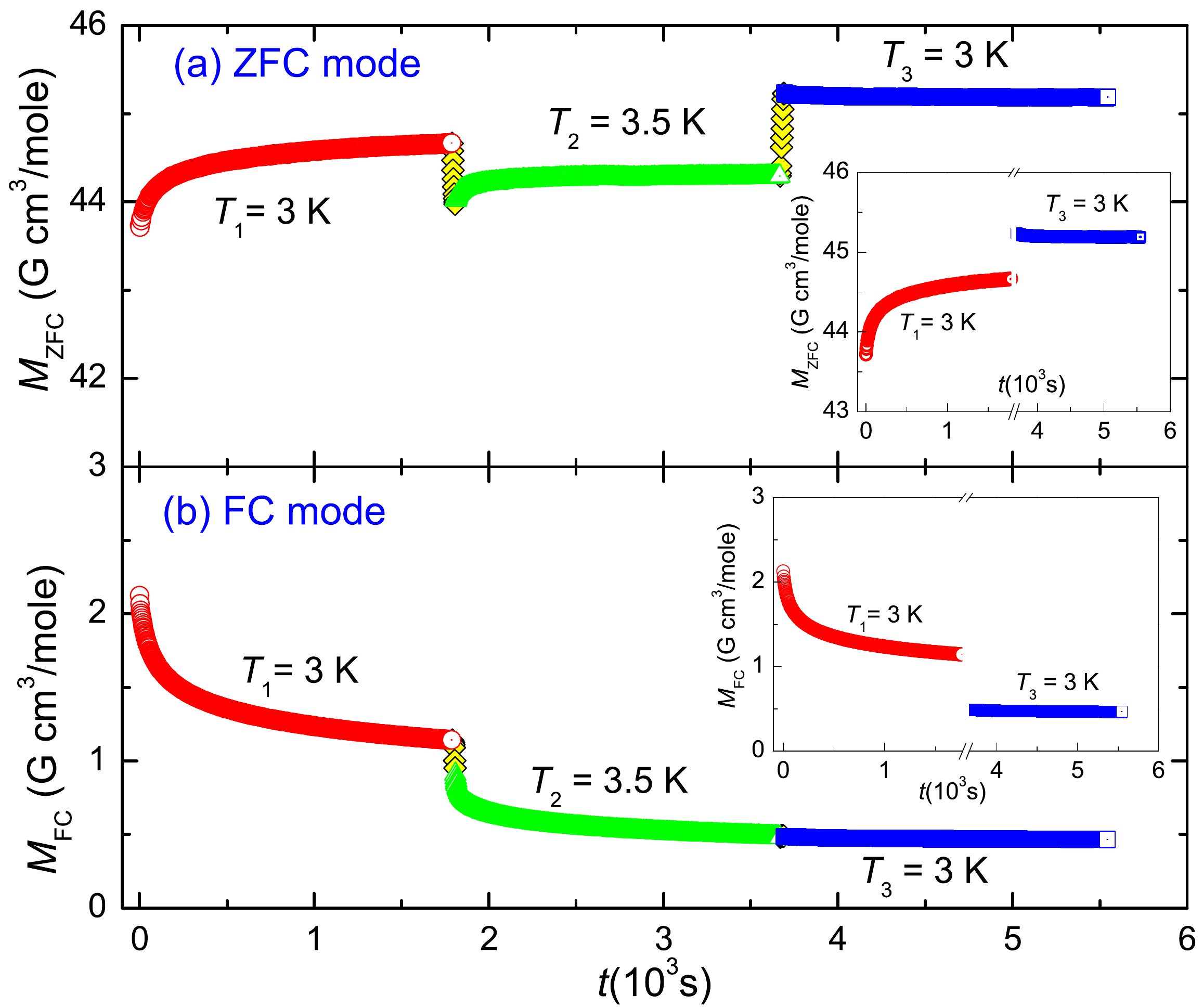}\caption{\label{fig:Positive cycle LRMO}{\small{}The magnetic relaxation in
LRMO with a positive heating cycle in (a) ZFC mode and (b) FC mode.
Insets show discontinuity in the relaxation data at $\mathrm{T}_{1}$
and $\mathrm{T}_{3}$ in both mode.}}
\end{figure}

To know further features of the memory effect, we have measured the
magnetic relaxation in ZFC and FC mode with a negative and a positive
temperature cycling as shown in Fig. \ref{fig:Negative cycle LRMO}
and Fig. \ref{fig:Positive cycle LRMO} respectively. We have recorded
each relaxation curve for $\frac{1}{2}$ hour. In the ZFC mode, we
have cooled the sample in the absence of field but measured the data
with an applied field of $\mathit{H}$ = 500\,Oe. In contrast to
that, in the FC mode, the field ($\mathit{H}$ = 500 Oe) is continuously
on during cooling of the sample and switched off just before the measurement
starts. In the negative heat cycle, we have quenched the system to
a lower temperature and resumed the relaxation process. The initial
and final relaxation data are at T$_{1}$ = T$_{3}$ = 3\,K whereas
the quenched one is at T$_{2}$ = 2.5\,K. Fig. \ref{fig:Negative cycle LRMO}
(a) and (b) shows the relaxation data in ZFC and FC mode in the negative
heat cycle process. If we ignore the middle one, the initial and the
final relaxation are just a continuation of each other as shown in
the inset of Fig. \ref{fig:Negative cycle LRMO} (a) and (b). So in
the negative heat cycle or the temporary quenching, the system remembers
the earlier states where it was, irrespective of the measurement processes
i.e. either ZFC or FC mode. This is the memory effect in negative
heating cycle.

We also have measured magnetization in a temporary heating cycle for  comparing the response with the negative heating cycle. In the positive
heat cycle, we have increased the temperature of the middle step to
T$_{2}$ = 3.5\,K whereas the initial and final steps are at T$_{1}$
= T$_{3}$ = 3\,K. Fig. \ref{fig:Positive cycle LRMO} (a) and (b)
show the relaxation data in ZFC and FC mode respectively. From the
inset of Fig. \ref{fig:Positive cycle LRMO}, it is very clear that
the relaxations at T$_{1}$ and T$_{3}$ are discontinuous and the
response of the system is asymmetric. So the positive heat cycle erases
the memory in both ZFC and FC processes. This supports the hierarchical
model as proposed for the spin-glasses.
\begin{center}
\textbf{4. Aging effect and relaxation}
\par\end{center}

\begin{figure}[h]
\begin{centering}
\includegraphics[scale=0.42]{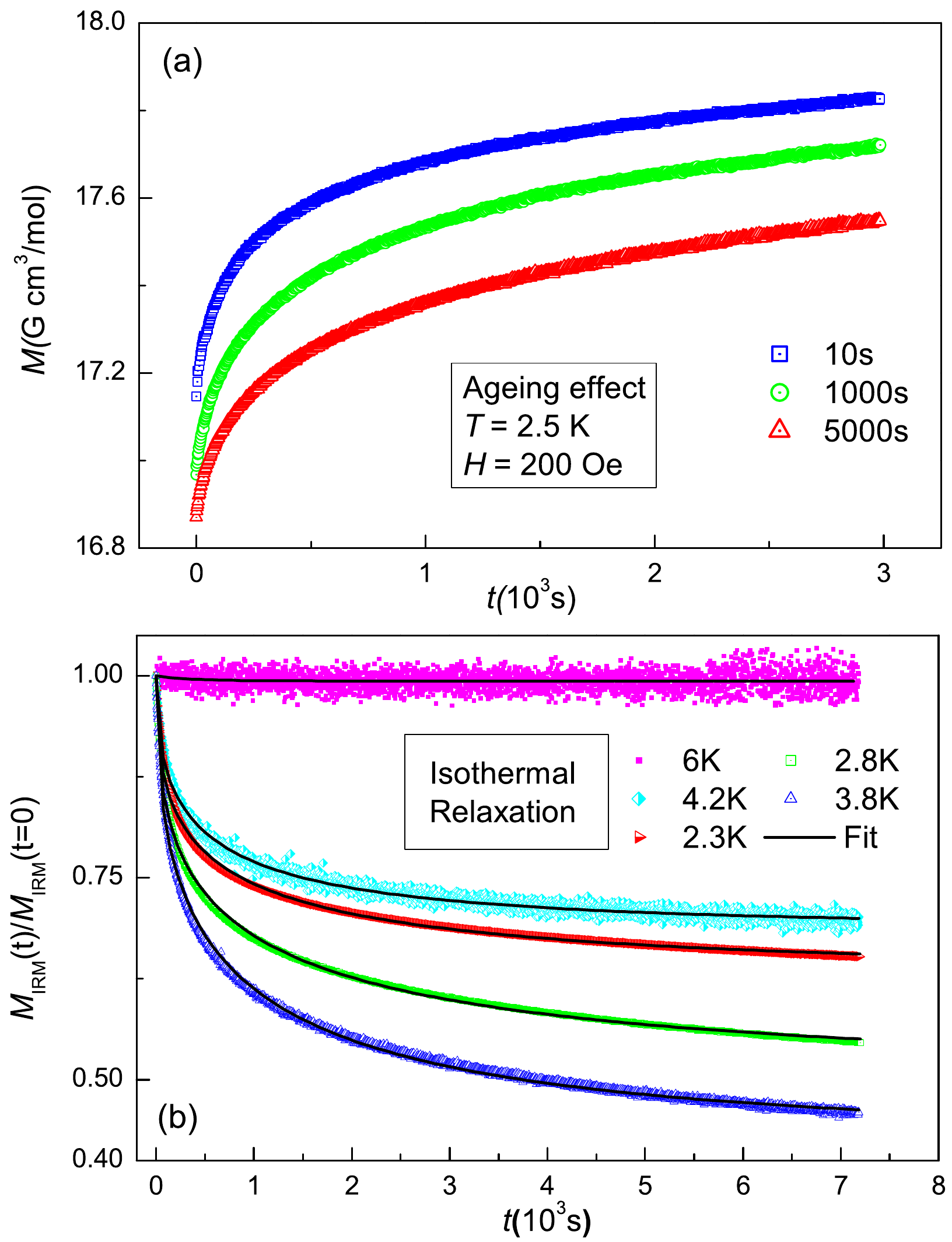}
\par\end{centering}
\caption{\label{fig:Aging-effect-of LRMO}{\small{}(a) Aging effect of the
dc magnetization of LRMO as a function of time with intervals of 10\,s,
1000\,s and 5000\,s.} \label{fig:Isothermal-remanent LRMO}{\small{}(b)
The normalized isothermal remanent magnetization is shown as a function
of time. The data are well fitted with a stretched exponential as
described in text.}}
\end{figure}

The Fig. \ref{fig:Aging-effect-of LRMO} shows the aging effect in
the dc magnetization data. Below the freezing temperature, that is
at 2.5\,K, the growth of the magnetization is recorded as a function
of time in the ZFC mode with an applied field of $H$ = 200\,Oe after
a different waiting time (t). We have waited for three different times
like 10\,s, 1000\,s and 5000\,s. From our plot, it is obvious that
the magnetization growth is faster for small waiting time and slower
for large waiting time. These point towards the formation of metastability of the glassy state.
We also measured the isothermal remanent magnetization ($M_{\mathrm{IRM}}$) i.e. the relaxation
of LRMO to probe the metastability further below
the spin-glass transition temperature. For this, a field of 300\,Oe
was applied for 300\,s after we cooled the LRMO sample in the zero
field mode and reached a particular temperature and then the applied
field was switched off and let the system to relax for 2 hours at
that temperature. During relaxation, the magnetization data was then
recorded as a function of time. Fig. \ref{fig:Isothermal-remanent LRMO}
shows the decay curves normalized to the magnetization before making
the field zero, $M_{\mathrm{IRM}}(t)/M_{\mathrm{IRM}}(0).$ These
isothermal remanent magnetization were well fitted with stretched
exponential $\left(M_{t}(H)=M_{0}(H)+[M_{\infty}(H)-M_{0}(H)][1-exp\{-(t/\tau)^{\alpha}\}]\right)$
and from the fitting, we obtained the characteristic relaxation time
$(\tau)$ at different temperatures (shown in Fig. \ref{fig:Isothermal-remanent LRMO}).
Here $\alpha$ is the stretching exponent (0$\leqslant\alpha\leqslant1),$
$\mathit{M_{\mathrm{0}}}$ and $\mathit{M_{\infty}}$ are the magnetization
at when $\mathit{t}$ $\rightarrow$ 0 and $\mathit{t}$ $\rightarrow$
$\infty$ respectively. The best fit parameters obtained for each
isotherm is listed in Table \ref{tab:Isotherm LRMO}. We do expect that the decay of $M_{\mathrm{IRM}}$ is faster for temperature closer
to $\mathrm{\mathit{T_{f}}}$. This signifies that the system forms
a metastable and irreversible state below $\mathrm{\mathit{T_{f}}}$.
As expected, above $\mathrm{\mathit{T_{f}}}$ i.e. at 6\,K, $M_{\mathrm{IRM}}$
is independent of time.

\begin{table}[H]
\centering{}\caption{\label{tab:Isotherm LRMO}{\small{}The best fit result of each isothermal
relaxation of LRMO fitted by stretched exponential.}}
\vspace{0.5cm}
\begin{tabular}{cccc}
\hline 
$\mathit{T}$(K) & Stretching exponent $\alpha$ & $\frac{\mathit{M_{\infty}}}{\mathit{M_{\mathrm{0}}}}$ & Relaxation time $\tau(\mathrm{s})$\tabularnewline
\hline 
2.3 & 0.42 & 0.63 & 638\tabularnewline
2.8 & 0.42 & 0.51 & 877\tabularnewline
3.8 & 0.45 & 0.43 & 770\tabularnewline
4.2 & 0.47 & 0.69 & 526\tabularnewline
\hline 
\end{tabular}
\end{table}

\begin{center}
\textbf{\large{}C. Heat capacity}{\large\par}
\par\end{center}

\begin{figure}[h]
\centering{}\includegraphics[scale=0.35]{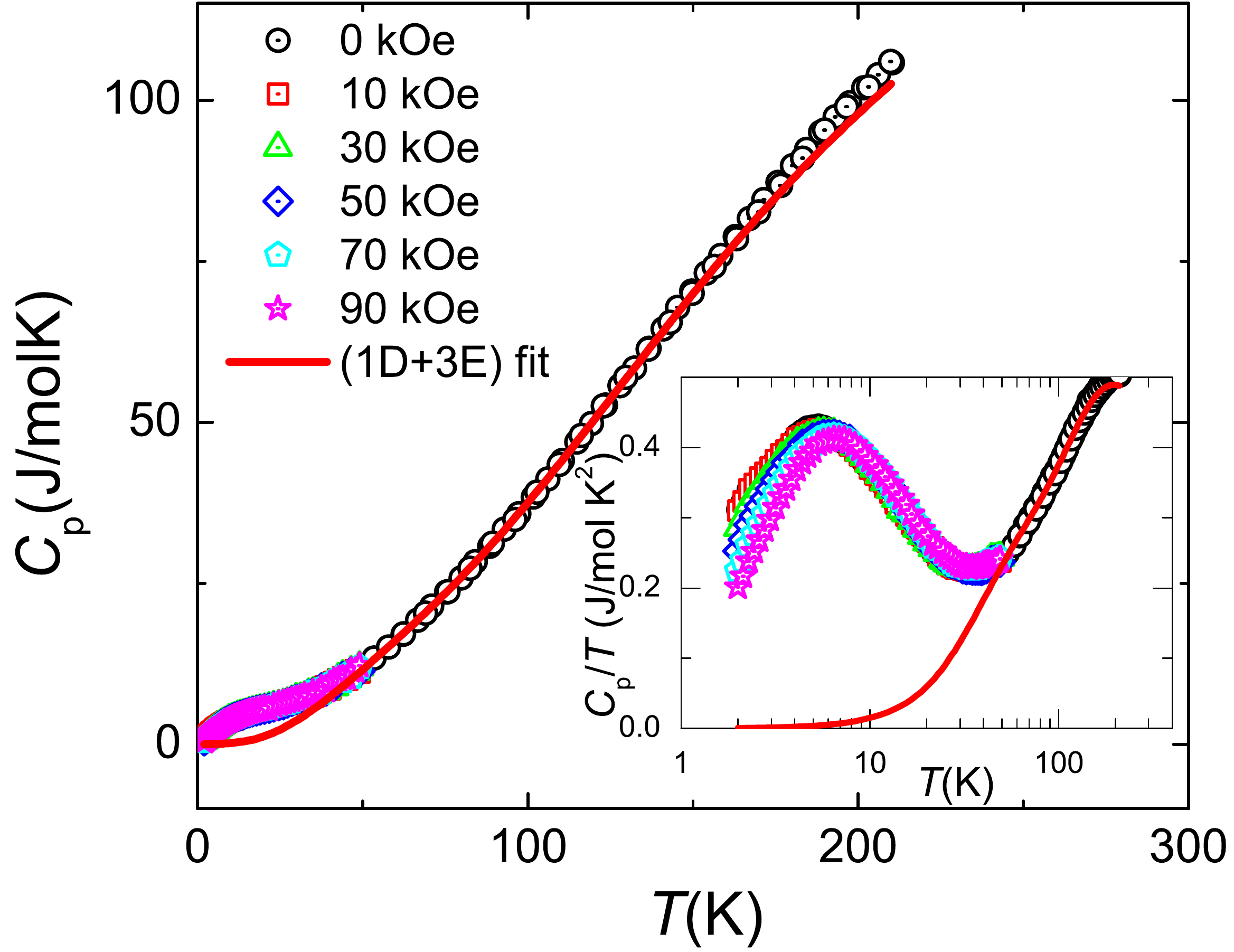}\caption{\label{fig:Total-heat-Capacity()}{\small{}Main figure shows the total
heat Capacity $C_{\mathrm{p}}(T)$ of $\mathrm{LRMO}$ measured at
different fields with Debye-Einstein fit (see text). In the inset
$C_{\mathrm{p}}(T)$/$\mathit{T}$ vs. $\mathit{T}$ plot shows the
prominent magnetic contribution to the $C_{\mathrm{p}}$.}}
\end{figure}

The heat capacity $C_{\mathrm{p}}(T)$ of magnetic LRMO was measured
at different fields (0 - 90\,kOe) in the temperature range 1.8- 300\,K.
There is no anomaly in the $C_{\mathrm{p}}(T)$ vs. $\mathit{T}$
data as might usually be expected for short-range order (SRO) or long-range order (LRO). In the inset
of Fig. \ref{fig:Total-heat-Capacity()}, $C_{\mathrm{p}}(T)/T$ vs.
$\mathit{T}$, there is no significant influence of the applied magnetic
field on the heat capacity. Also, no Schottky type anomaly was found
in this system at the low temperature.

The total heat capacity of LRMO has the contribution from lattice
($\mathit{C_{\mathrm{lat}}}$) and magnetic ($\mathit{C_{\mathrm{m}}}$)
both. As there was no suitable non-magnetic analogue available we
have fitted the data with Debye term $\left(C_{\mathrm{d}}\left[9nR(\frac{T}{\theta_{D}})^{3}\intop_{0}^{x_{D}}\frac{x^{4}e^{x}}{(e^{x}-1)^{2}}dx\right]\right)$
and several Einstein terms $\left(\sum C_{\mathrm{e}_{i}}\left[3nR(\frac{\theta_{E_{i}}}{T})^{2}\frac{exp(\frac{\theta_{E_{i}}}{T})}{(exp(\frac{\theta_{E_{i}}}{T})-1)^{2}}\right]\right)$
in the $\mathit{T}$-range (55-130)\,K and then extrapolated to low-$\mathit{T}$
to determine the $\mathit{C_{\mathrm{lat}}}$. Out of them, one Debye
term plus two Einstein terms (1D+2E) fit is the best one where
the coefficients $C_{\mathrm{d}}$ and $C_{\mathrm{e_{i}}}$ accounts
for the relative weight of the acoustic and optical modes of vibrations
respectively. After fitting we obtained $C_{\mathrm{d}}$:$C_{\mathrm{e_{1}}}$:$C_{\mathrm{e_{2}}}$
= 1:1:5. The deviation of the $C_{\mathrm{p}}(T)$
from the Debye-Einstein fit below 50\,K indicates the presence of
a significant magnetic contribution to the heat capacity. The magnetic
heat capacity $C_{\mathrm{m}}(T)$ is obtained by subtracting the
lattice contribution from total heat capacity and shown in Fig. \ref{fig:(a)Magnetic-Heat-capacity_LRMO}
on the left $\mathit{y}$-axis. The magnetic heat capacity is almost
independent of the strength of the applied field. It shows a hump
around 18\,K which indicates onsets of short-range interactions among
the magnetic atoms. Also the magnetic entropy change $\Delta S_{\mathrm{m}}(T)$
is calculated using $\Delta S_{\mathrm{m}}=\int\frac{C_{\mathrm{m}}}{T}dT$
relation and shown in the right $\mathit{y}$-axis of Fig. \ref{fig:(a)Magnetic-Heat-capacity_LRMO}.
Its value is 8.55\,(J/mol K) which is 75\% of the expected 11.52\,(J/mol
K) for $\mathit{S}$ = $\frac{3}{2}$ spin. Considering the uncertainty
involved in determining the lattice specific heat, the value of $\Delta S_{\mathrm{m}}$
obtained is not far from the expected value.
\begin{center}
\begin{figure}[h]
\centering{}\includegraphics[scale=0.3]{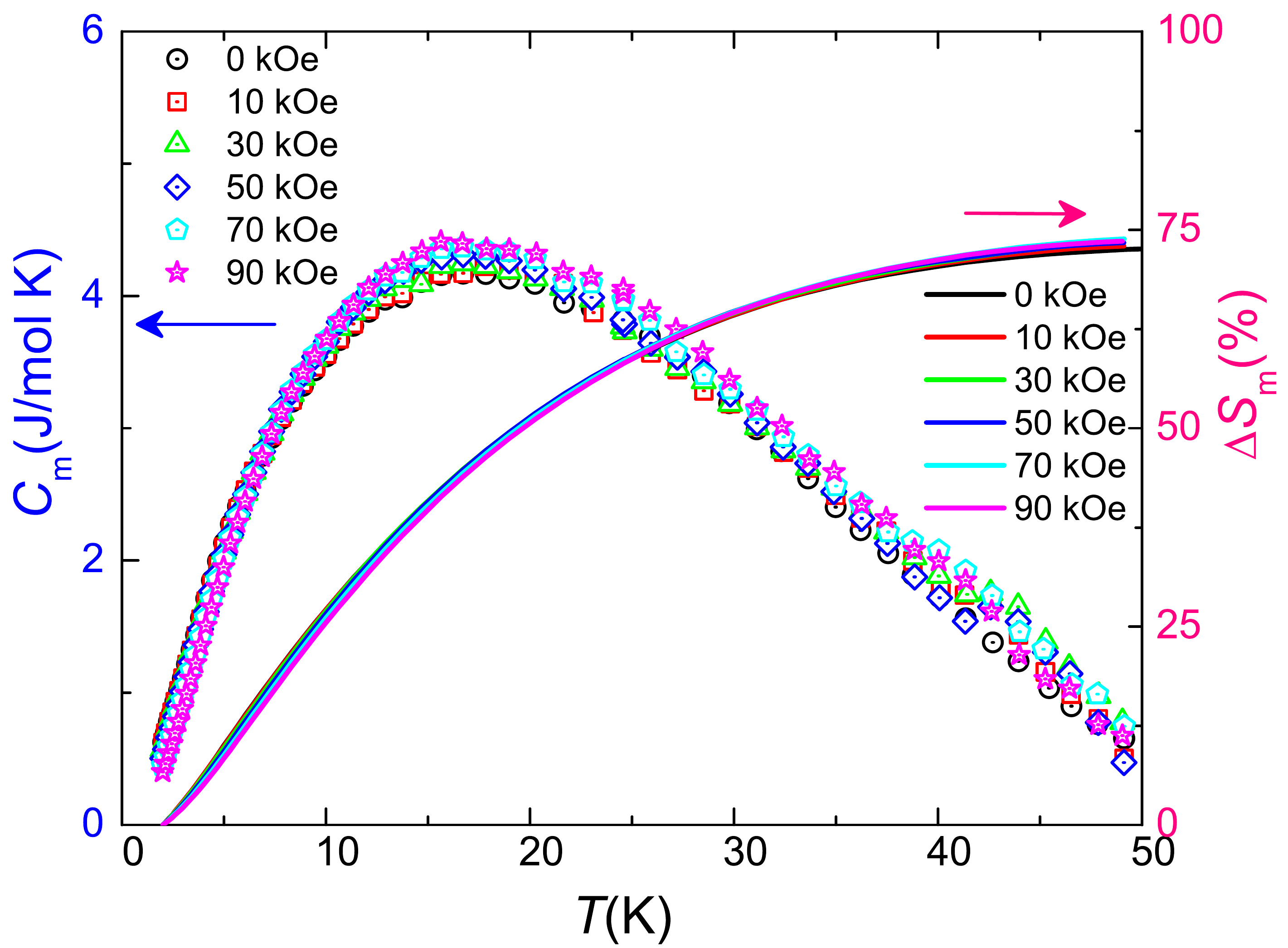}\caption{\label{fig:(a)Magnetic-Heat-capacity_LRMO}{\small{}The magnetic heat
capacity $C_{\mathrm{m}}(T)$ on left $\mathit{y}$-axis and the percentage of magnetic entropy change $\Delta S_{\mathrm{m}}(T)$ on the right $\mathit{y}$-axis
of LRMO sample are shown.}}
\end{figure}
\par\end{center}

\begin{center}
\textbf{\large{}D. $\mathrm{^{7}Li}$ NMR Result}{\large\par}
\par\end{center}

$\mathrm{{\normalcolor ^{7}Li}}$ nuclei has a high natural abundance
(92.6\%) and it has nuclear spin $\mathit{I}$ = $\frac{3}{2}$ with
the value of gyromagnetic ratio $\frac{\gamma}{2\pi}$ = 16.54607\,(MHz/T).
We have measured the field sweep $\mathrm{^{7}Li}$ NMR at 60\,MHz
and 95\,MHz. We also measured spin-lattice relaxation rate (1/$T_{1}$)
and the spin-spin relaxation rate (1/$T_{2}$) at 60\,MHz ($\mathit{H}$$\sim$36\,kOe).
These measurements throw light on the nature of the intrinsic interactions
of magnetic atoms.
\begin{center}
\textbf{1. $\mathrm{^{7}Li}$ NMR Spectra}
\par\end{center}

\begin{figure}[h]
\centering{}\includegraphics[scale=0.35]{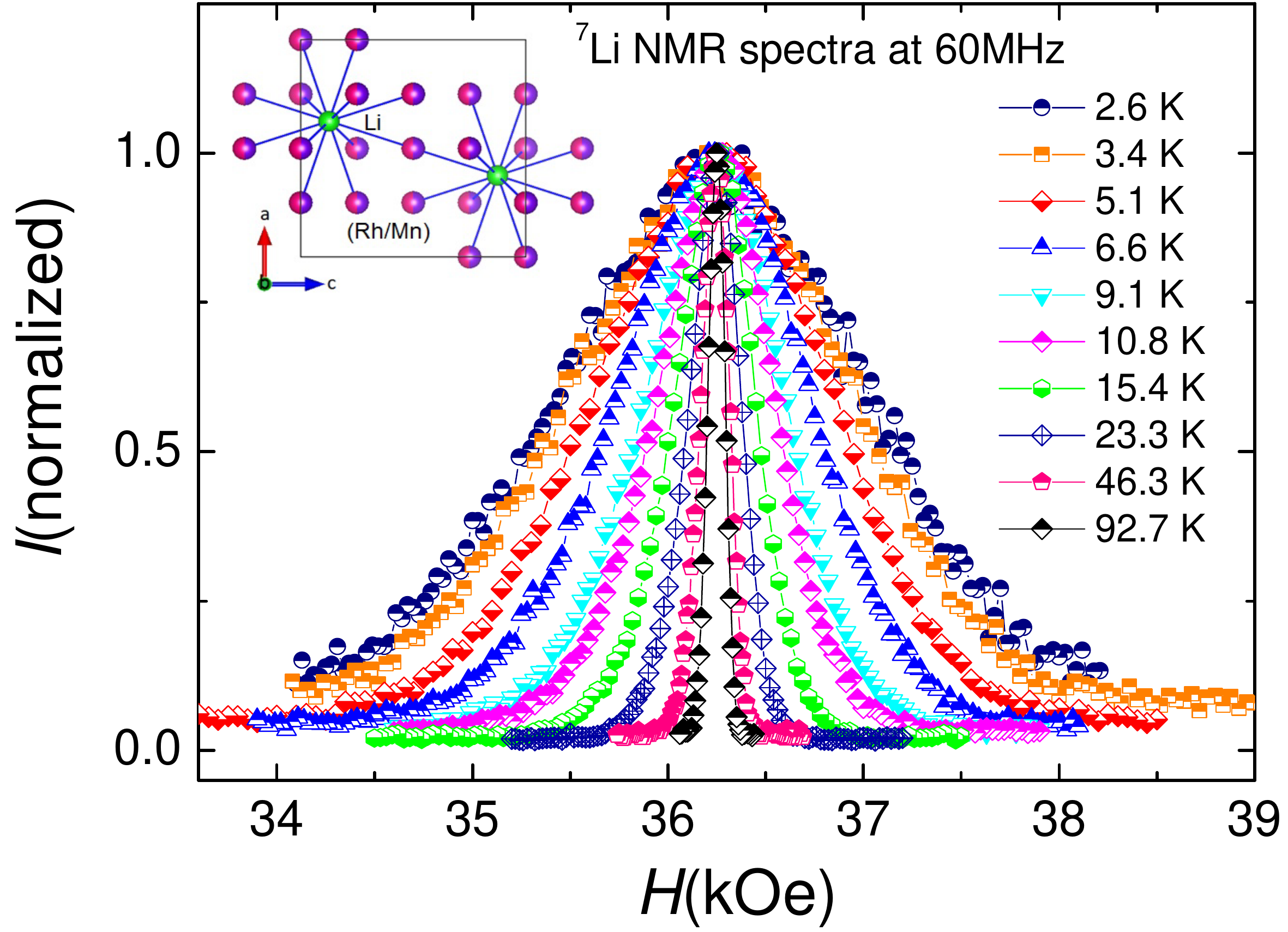}\caption{\label{fig:NMR-spectra-ofLRMO}{\small{}Normalized $\mathrm{\mathrm{^{7}L}i}$
NMR spectra of $\mathrm{LRMO}$ measured at 60\,MHz frequency and
at different temperatures down to 2.6\,K.}}
\end{figure}

\begin{figure}[h]

\centering{}\includegraphics[scale=0.35]{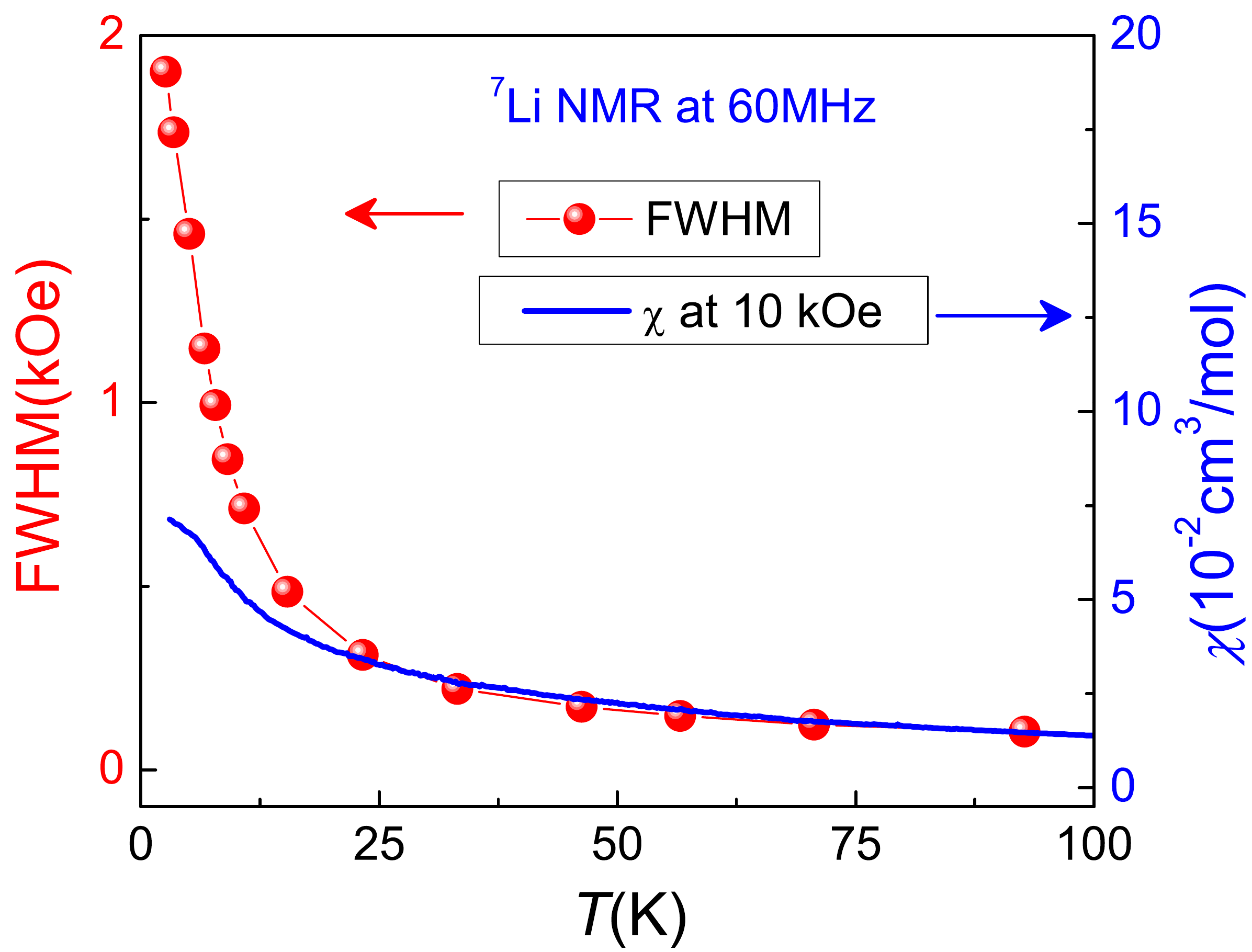}\caption{\label{fig:FWHM-follows-dc susceptiblity}{\small{}The FWHM (shown
on left $\mathit{y}$-axis) follows the bulk dc susceptibility (on
right $\mathit{y}$-axis) at high-$\mathit{T}$.}}
\end{figure}

For the spectra, we use the optimal pulse sequence ($\pi/2-\tau-\pi$)
= 5\,$\mathrm{\mu s}$-100\,$\mathrm{\mu s}$-10\,$\mathrm{\mu s}$
at 60\,MHz. The 100\,$\mathrm{\mu s}$ refers to the time duration
between the starting of the two pulses. The spectra at different temperatures
(from 92.7\,K to 2.6\,K) are shown in Fig. \ref{fig:NMR-spectra-ofLRMO}.
There is no significant shift in the spectra. The lithium surroundings
in one unit cell are shown in the inset of Fig. \ref{fig:NMR-spectra-ofLRMO}.
From the spectra, we have obtained the full width at half maxima (FWHM)
at different temperatures which track the dc susceptibility well as
shown in Fig. \ref{fig:FWHM-follows-dc susceptiblity}.

\begin{figure}[h]
\centering{}\includegraphics[scale=0.33]{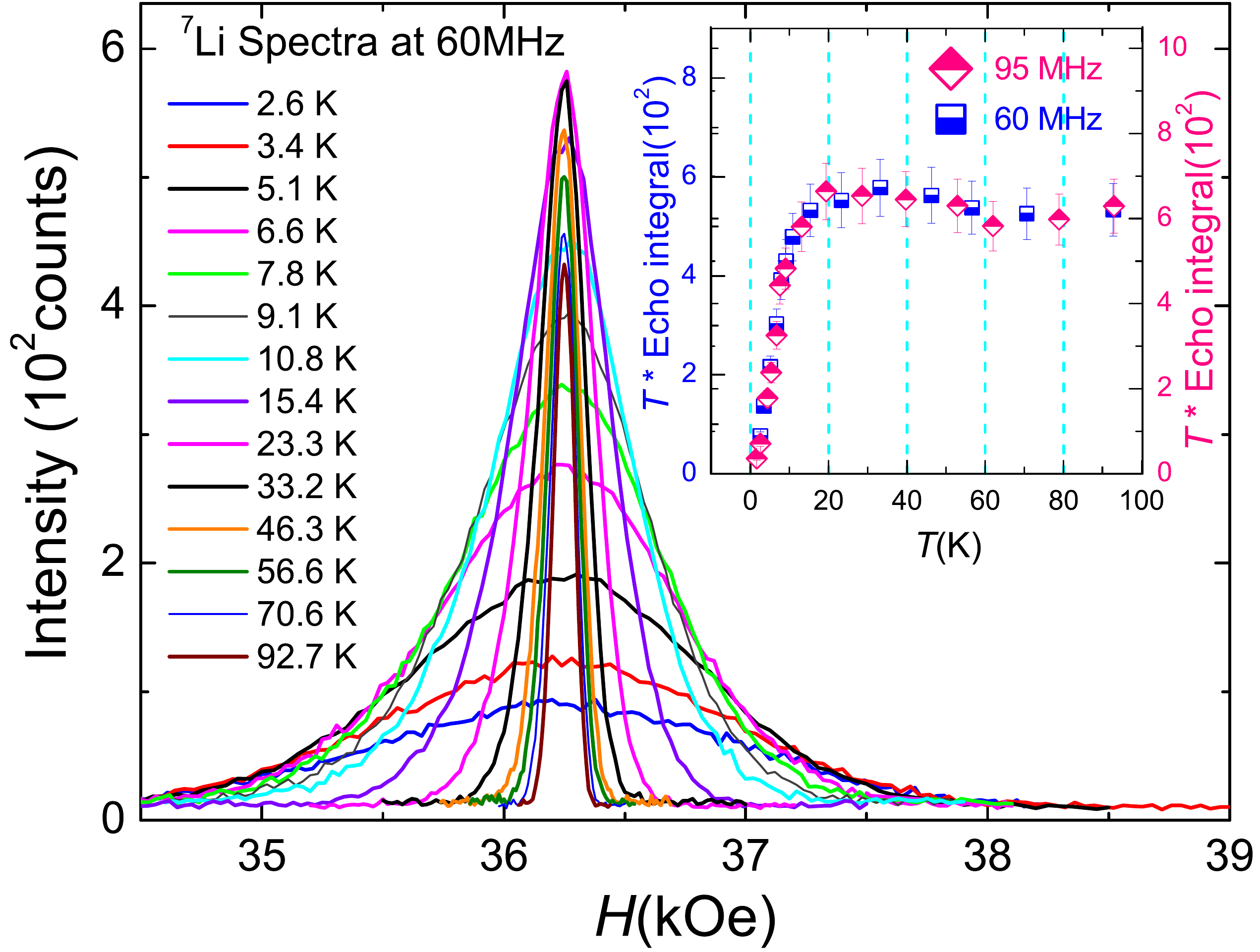}\caption{\label{fig:LRMO_Int times T}{\small{}The $\mathrm{^{7}Li}$ Spectra
at different temperatures as a function of field $\mathit{H}$(T),
In the inset the corrected echo integral intensity times the temperature
as a function of temperature.}}
\end{figure}

The $\mathrm{^{7}Li}$ spectra at different temperatures are plotted
without normalization of the spin-echo intensity $\mathit{\mathrm{\mathit{I}}}$
as a function of sweep field $\mathit{H}$ (see Fig. \ref{fig:LRMO_Int times T}).
On lowering the temperature the total spectral intensity is constant
down to about 20\,K and then begins to decrease below 20\,K. In
the inset of Fig. \ref{fig:LRMO_Int times T}, the echo integral (which
is obtained by integrating the line-shape as a function of field $\mathit{i.e.}$
the area under the spectrum at a particular temperature) times the
temperature is plotted as a function of temperature. It shows a drop
below 20\,K. This suggests a loss of signal most probably due to
development of frozen magnetic regions in the sample below 20\,K.
\begin{center}
\textbf{2. Spin-lattice relaxation rate, 1/$\mathit{T_{\mathrm{1}}}$}
\par\end{center}

The spin-lattice relaxation rate $(1/T_{1}$) of $^{7}\mathrm{Li}$
was measured by using a saturation recovery of the longitudinal magnetization
using saturation pulse ($\mathit{\frac{\pi}{2}\mid_{\mathrm{sat}}}$)
of 10\,$\mathrm{\mu s}$ at various temperatures from 93\,K to 2.65\,K.
The saturation recovery curves are shown in the inset of Fig. \ref{fig:Saturation-recovery-of 7Li_LRMO}.
The curve above 7\,K are best fitted with the single exponential
function $[\mathrm{1-\frac{\mathit{M}(t)}{\mathit{M}(0)}=\mathit{A}\thinspace exp^{(\frac{-t}{\mathit{T}_{1}})}}]$
and below about 7\,K are best fitted with a stretched exponential
function {[}$\mathrm{1-\frac{\mathit{M}(t)}{\mathit{M}(0)}=\mathit{B}\thinspace exp^{(\frac{-t}{\mathit{T}})^{\beta}}+\mathit{C}}${]}.
Here $\mathit{A\,\mathrm{\&}\,B}$ denotes the saturation level of
the signal and $\beta$ is the stretching exponent. In general, the
spin-glass systems possess a distribution of the spin-lattice relaxation
times ($\mathit{T_{\mathrm{1}}}$) due to the existence of different
relaxation channels. That's why here, $\beta$ determines the width
of the distribution window. This stretched exponential characteristics
of the saturation recovery data below about 7\,K confirms the presence
of discrete and local magnetic domains. The Fig. \ref{fig:Saturation-recovery-of 7Li_LRMO}
shows the spin-lattice relaxation rate as a function of temperature.
Below 20\,K it starts to increase and at around 7\,K it shows a
peak. It appears that the onset of freezing of the magnetic regions
starts around 20\,K and at 7\,K they lock into a spin-glass state.
This supports the dc magnetic susceptibility as well as the magnetic
heat capacity data which shows a hump just below 20\,K.
\begin{center}
\begin{figure}[h]
\begin{centering}
\includegraphics[scale=0.4]{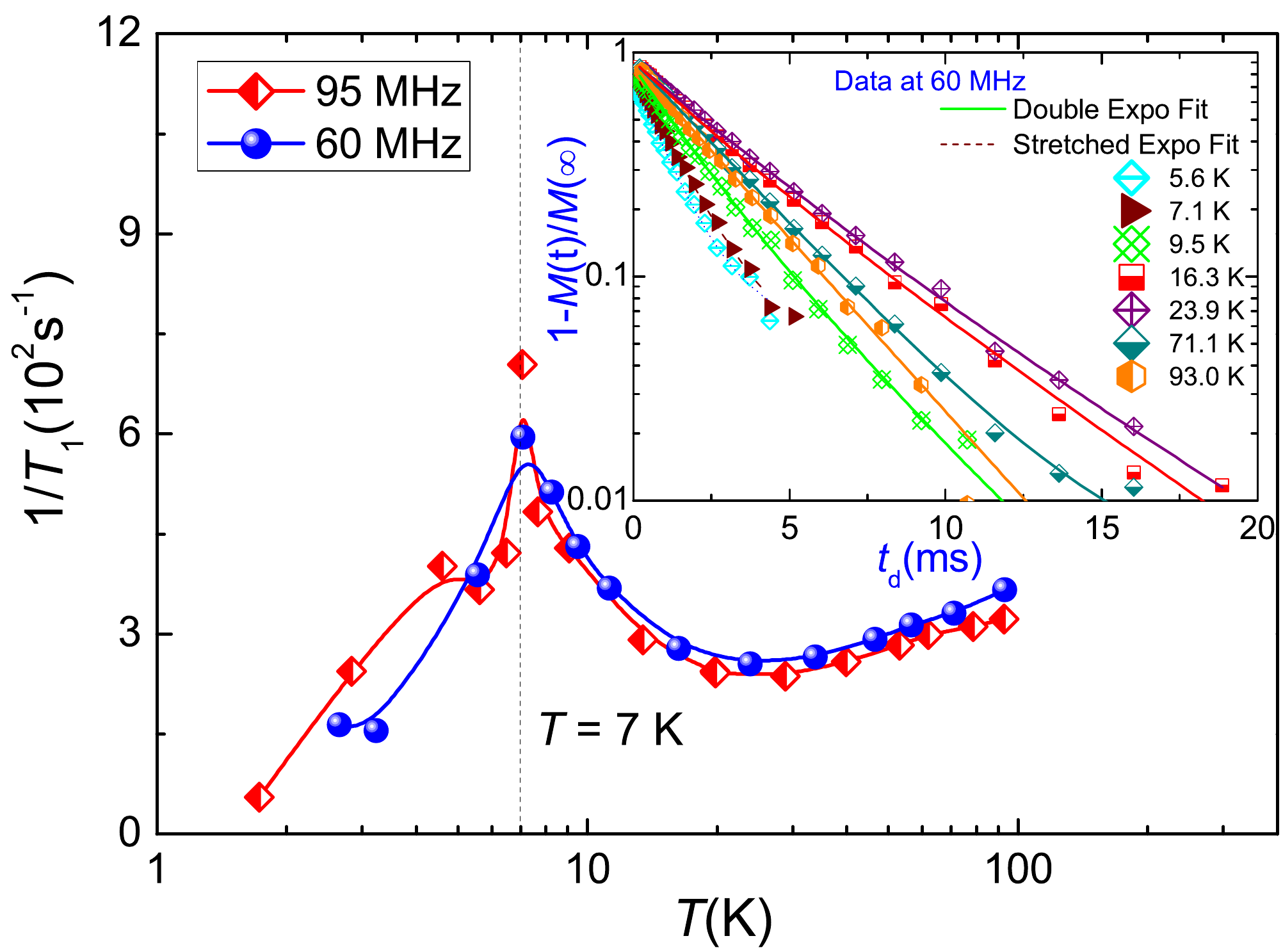}
\par\end{centering}
\caption{\label{fig:Saturation-recovery-of 7Li_LRMO}{\small{} The spin-lattice
relaxation rate as a function of temperatures at two frequencies 60\,MHz
and 95\,MHz show ordering at 7\,K. In the inset, the saturation
recovery of the longitudinal magnetization as a function of the delay
time at various temperatures. The solid and dashed lines are the best
fit for single exponential function (above 7\,K) and stretched exponential
function (below 7\,K) respectively.}}

\end{figure}
\textbf{3. Spin-spin relaxation rate, 1/$\mathit{T_{\mathrm{2}}}$}
\par\end{center}

\begin{figure}[h]
\begin{centering}
\includegraphics[scale=0.4]{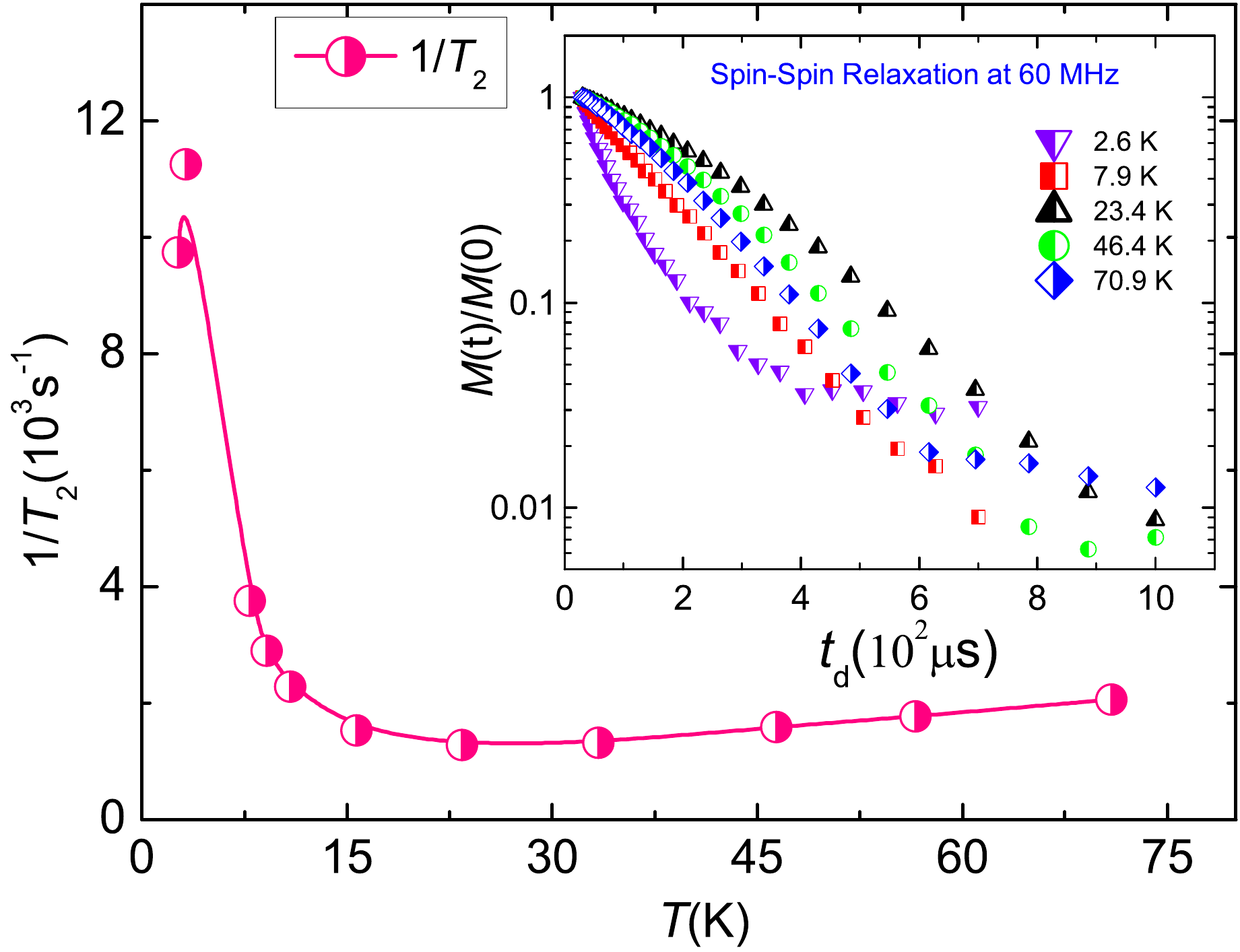}\caption{\label{fig:Saturation-recovery-of spin-spin_LRMO}{\small{}The temperature
variation of the spin-spin relaxation rate 1/$\mathit{T_{\mathrm{2}}}$.
The inset shows the decay of the transverse nuclear magnetization
at some selected temperatures.}}
\par\end{centering}
\end{figure}

The inset of Fig. \ref{fig:Saturation-recovery-of spin-spin_LRMO}
shows the decay of the transverse nuclear magnetization data at 60
MHz with different temperatures. The data above 7\,K are well fitted
to a Gaussian modified exponential function {[}$\mathrm{\frac{\mathit{M}(t)}{\mathit{M}(0)}=\mathit{M}(0).\{exp^{-(\frac{t}{\mathit{T}_{2G}})^{2}}.exp^{(\frac{-2t}{\mathit{T}_{2}})}}\}${]}
and the data below 7\,K are fitted with a stretched exponential function
{[}$\frac{M(t)}{M(0)}=A_{2}exp{}^{(\frac{-2t}{T_{2}})^{\beta}}+C${]}.
From fitting, we obtained the $T_{\mathrm{2}}$ values and plotted
the spin-spin relaxation rate (1/$\mathit{T_{\mathrm{2}}}$) as a
function of temperature in Fig. \ref{fig:Saturation-recovery-of spin-spin_LRMO}.
It shows that the spin-spin correlation begins to increase around
15\,K with a peak at $\thickapprox$ 3.2\,K.

\section{conclusions}

With respect to the crystallography of polycrystalline LRMO we confirmed
a single-phase nature from our XRD investigation. In $\chi$($\mathit{T}$),
ZFC-FC bifurcation was found below 4.45\,K which is very much field
sensitive. This ZFC-FC splitting suggests the presence of a glassy
state. The frequency dependent $\chi_{\mathrm{ac}}$, where the freezing
temperature ($\mathrm{\mathit{T_{f}}}$) shifts towards higher $\mathit{T}$
values as the frequency increases is a signature of glassy systems
and thus it confirms the presence of the spin-glass ground state.
Also, the out of phase component of the ac susceptibility $\chi_{\mathrm{ac}}^{''}(T)$
has a frequency dependence with an anomaly around $\mathit{T_{f}}$.
The $\chi_{\mathrm{ac}}^{''}(T)$ is non-zero positive below $\mathit{\mathit{\mathit{\mathit{\mathit{T_{f}}}}}}$
and is negative above $\mathit{T_{f}}$. This observation ruled out
any bond disordered antiferromagnetic state. The characteristic frequency
$\omega_{0}\thinspace\approx$ $1.01\times10^{8}$\,rad/s obtained
from the Vogel \textendash Fulcher fit is less than that of conventional
spin-glass systems $(10^{13}$) \,rad/s, but the characteristic time
$\tau_{0}$ $\approx$ 2.85 $\times$10$^{-10}$\,s and critical
exponent $\mathit{zv}$ $\approx$ 4.88 values are close to a conventional
spin-glass \cite{Luo2008}. This implies that the ground state of
LRMO is more likely to be a conventional spin-glass. From heat capacity
measurement, there occurs significant contribution of magnetic heat
capacity and no sharp anomaly presents down to 2\,K. The calculated
magnetic entropy change $\Delta S_{\mathrm{m}}=8.55\,(\mathrm{J/mol\thinspace K})$
is 75\% of the theoretical value $\mathit{R}$ln(4) for this system.
These numbers are not far from the usual LRO transition. However the
change of entropy starts to decrease below 30\,K, which is close
to the CW temperature $\mid\theta_{\mathrm{CW}}\mid=26\thinspace\mathrm{K}$
also. From $\mathrm{^{7}Li}$ NMR, there is no significant shift of
the spectrum and the FWHM of spectra at high temperatures follows
the Curie-Weiss behavior like dc susceptibility. The echo integral
intensity times the $\mathit{T}$ vs. $\mathit{T}$ shows a drop below
20\,K. This suggests a loss of signal probably due to development
of frozen magnetic domains within the sample. In order to shed more
light on the spin dynamics of $\mathrm{Mn^{4+}}$ ions, we have measured
spin-lattice relaxation rate ($1/T_{1}$) and spin-spin relaxation
rate (1/$T_{2}$) for $\mathrm{^{7}Li}$ nuclei. Both show anomalies
below 7\,K like in the dc susceptibility indicating the spin-glass
ground state of $\mathrm{LRMO}$.

\section{acknowledgement}

SK acknowledges the discussion with Dr. Aga Shahee and R. K. Sharma
and the financial support from IRCC, IIT Bombay. AVM would like to
thank the Alexander von Humboldt foundation for financial support
during his stay at Augsburg Germany. We kindly acknowledge support
from the German Research Society (DFG) via TRR80 (Augsburg, Munich).\bibliographystyle{apsrev4-1}
\bibliography{Citation}

\end{document}